\documentclass[aps,preprint, prl, superscriptaddress, showpacs,nobibnotes]{revtex4-1}
\usepackage{graphicx}
\usepackage{bm,amsmath}
\usepackage{dcolumn}
\usepackage{natbib,xcolor}
\usepackage{esvect}

\begin{document}
\title{Massive Dirac fermions in a ferromagnetic kagome metal}

\author{Linda Ye}
\thanks{These authors contributed equally}
\affiliation{Department of Physics, Massachusetts Institute of Technology, Cambridge, Massachusetts 02139, USA}
\author{Mingu Kang}
\thanks{These authors contributed equally}
\affiliation{Department of Physics, Massachusetts Institute of Technology, Cambridge, Massachusetts 02139, USA}
\author{Junwei Liu}
\altaffiliation{Current Affiliation: Department of Physics, Hong Kong UST, Clear Water Bay, Hong Kong, China}
\affiliation{Department of Physics, Massachusetts Institute of Technology, Cambridge, Massachusetts 02139, USA}
\author{Felix von Cube}
\altaffiliation{Current Affiliation: Hitachi High-Technologies Europe GmbH, Krefeld, Germany}
\affiliation{Harvard John A. Paulson School of Engineering and Applied Sciences, Harvard University, Cambridge, Massachusetts 02138, USA}
\author{Christina R. Wicker}
\affiliation{Department of Physics, Massachusetts Institute of Technology, Cambridge, Massachusetts 02139, USA}
\author{Takehito Suzuki}
\affiliation{Department of Physics, Massachusetts Institute of Technology, Cambridge, Massachusetts 02139, USA}
\author{Chris Jozwiak}
\affiliation{Advanced Light Source, E. O. Lawrence Berkeley National Laboratory, Berkeley, California 94720, USA}
\author{Aaron Bostwick}
\affiliation{Advanced Light Source, E. O. Lawrence Berkeley National Laboratory, Berkeley, California 94720, USA}
\author{Eli Rotenberg}
\affiliation{Advanced Light Source, E. O. Lawrence Berkeley National Laboratory, Berkeley, California 94720, USA}
\author{David C. Bell}
\affiliation{Harvard John A. Paulson School of Engineering and Applied Sciences, Harvard University, Cambridge, Massachusetts 02138, USA}
\affiliation{Center for Nanoscale Systems, Harvard University, Cambridge, Massachusetts 02138, USA}
\author{Liang Fu}
\affiliation{Department of Physics, Massachusetts Institute of Technology, Cambridge, Massachusetts 02139, USA}
\author{Riccardo Comin}
\thanks{Correspondence should be addressed to rcomin@mit.edu or checkelsky@mit.edu}
\affiliation{Department of Physics, Massachusetts Institute of Technology, Cambridge, Massachusetts 02139, USA}
\author{Joseph G. Checkelsky}
\thanks{Correspondence should be addressed to rcomin@mit.edu or checkelsky@mit.edu}
\affiliation{Department of Physics, Massachusetts Institute of Technology, Cambridge, Massachusetts 02139, USA}
\date{\today}

\maketitle

\textbf{The kagome lattice is a two-dimensional network of corner-sharing triangles \cite{CrystalStructure} known as a platform for exotic quantum magnetic states \cite{KagomeTheory, jarosite, volborthite, herber, CuBDC, QSL-RMP}.  Theoretical work has predicted that the kagome lattice may also host Dirac electronic states \cite{Mazin} that could lead to topological \cite{kagomeTI} and Chern \cite{kagomeCI} insulating phases, but these have evaded experimental detection to date.  Here we study the $d$-electron kagome metal Fe$_{3}$Sn$_{2}$ designed to support bulk massive Dirac fermions in the presence of ferromagnetic order.  We observe a temperature independent intrinsic anomalous Hall conductivity persisting above room temperature suggestive of prominent Berry curvature from the time-reversal breaking electronic bands of the kagome plane.  Using angle-resolved photoemission, we discover a pair of quasi-2D Dirac cones near the Fermi level with a 30 meV mass gap that accounts for the Berry curvature-induced Hall conductivity.  We show this behavior is a consequence of the underlying symmetry properties of the bilayer kagome lattice in the ferromagnetic state with atomic spin-orbit coupling.  This report provides the first evidence for a ferromagnetic kagome metal and an example of emergent topological electronic properties in a correlated electron system.  This offers insight into recent discoveries of exotic electronic behavior in kagome lattice antiferromagnets \cite{Mn3Sn, Felser, Kuroda} and may provide a stepping stone toward lattice model realizations of fractional topological quantum states \cite{kagomeFCI, kagomeTSM}.}

\begin{figure}
\includegraphics[width =  \columnwidth]{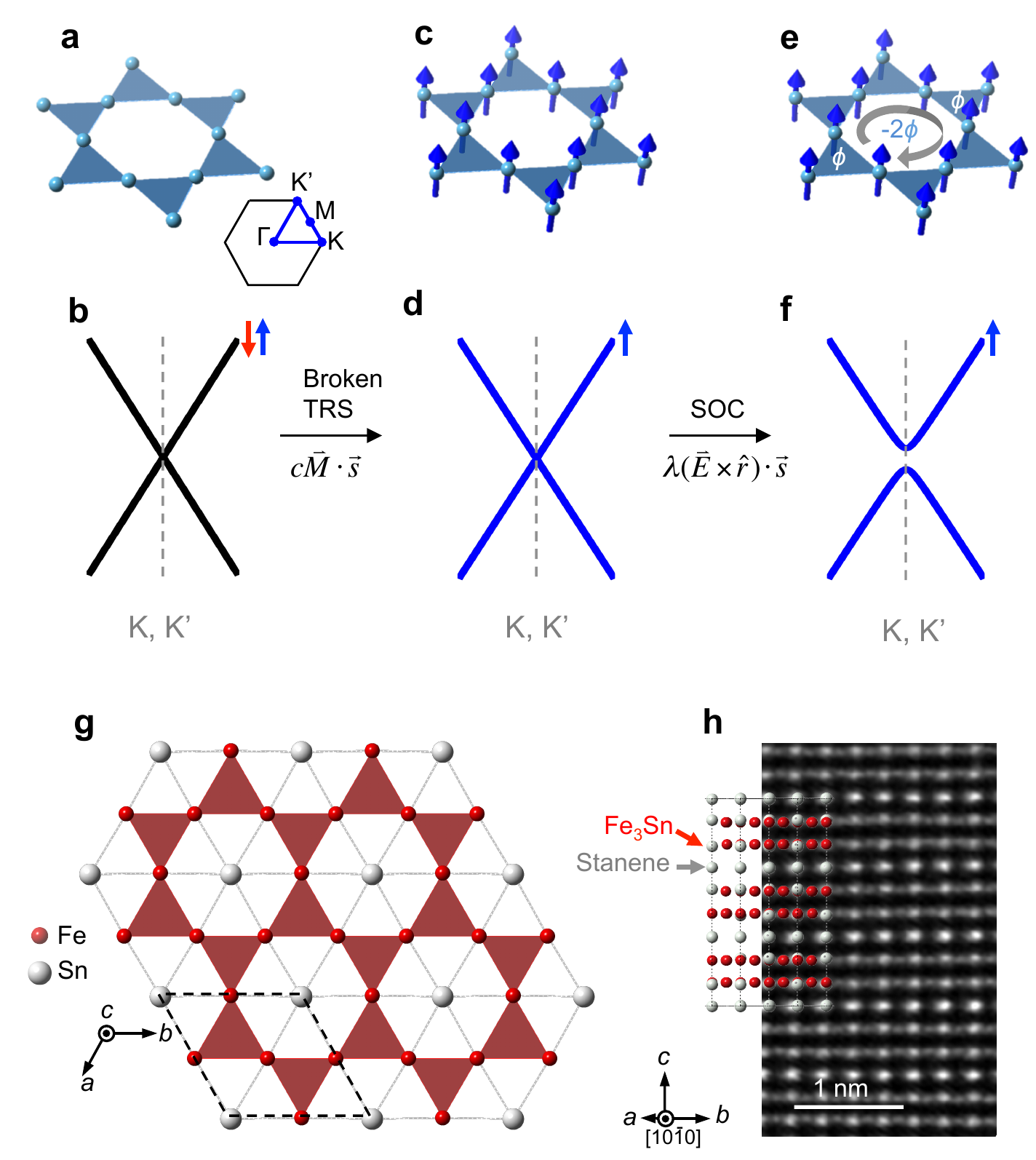}
\caption{\label{fig1} Ye \emph{et al.}}
\end{figure}

The kagome lattice (Fig. \ref{fig1}a) is a network with the $(3.6)^{2}$ Archimedes tiling predicted to support a wide variety of exotic phenomena in solids including magnetic frustration \cite{QSL-RMP}, flat band ferrromagnetism \cite{flatbandF}, and exotic superconductivity \cite{Superconductivity}.  Among these, significant interest has been focused on the frustration induced quantum spin liquid phase where the kagome lattice has played an important role in both theoretical and experimental progress \cite{KagomeTheory, jarosite, volborthite, herber}.  In terms of electronic structure, simple tight-binding models on the kagome lattice have long been known to yield unusual features including dispersionless (flat) bands and Dirac points (Fig. \ref{fig1}b), the latter in a manner similar to the hexagonal graphene lattice \cite{grapheneband}.  While such features have not been heretofore observed in experiment, theoretical interest has persisted and lead to several further predictions.  Of particular interest are kagome networks in which time reversal symmetry (TRS) is broken via ferromagnetism (Fig. \ref{fig1}c) \cite{Mazin, kagomeCI, kagomeFCI}, which has the effect of splitting the spin-degenerate Dirac bands (Fig. \ref{fig1}d).  Further inclusion of spin-orbit coupling (Fig. \ref{fig1}e) yields a variety of gapped phases (Fig. \ref{fig1}f) with possible integer \cite{kagomeCI, LiangCI} or fractional \cite{kagomeFCI, kagomeTSM} Chern numbers.  Such systems with exchange split bands and spin-orbit coupling in principle afford detection of their associated Berry curvature via Hall currents \cite{AHEHaldane}.

Despite the rich theoretical understanding that exists for electronic Berry phase effects in ferromagnetic kagome models, experimental realization of such behavior has been challenging in part owing to the relative rarity of the structure.  Kagome magnetic systems have been successfully studied in electrically insulating compounds such as jarosite \cite{jarosite}, volborthite \cite{volborthite}, and herbertsmithite \cite{herber}. Insulating ferromagnets of this type such as Lu$_2$V$_2$O$_7$ \cite{magnonHall} and Cu(1,3-bdc) \cite{CuBDC} have further shown connections to Berry curvature effects in the magnetic sector.  An approach to realizing instead a metallic kagome network has been reported in the hexagonal transition metal stannides $A_x$Sn$_y$ ($A=$ Mn, Fe, Co, $x:y=3:1,$ $3:2,$ $1:1$) \cite{AllFeSn}.  As shown in Fig. \ref{fig1}g for $A=$ Fe (studied here), starting from a single layer of the hexagonal closed packing structure of Fe, a kagome net naturally emerges by replacing a $2\times2$ sublattice (dashed cell) with Sn, resulting in an Fe$_{3}$Sn plane with an underlying Fe kagome lattice.  

We study here the bilayer kagome compound Fe$_{3}$Sn$_{2}$ (space group $R\bar{3}m$ with hexagonal lattice constants $a=5.338$ \AA\ and $c=19.789$ \AA), a structural variation of Fe$_{x}$Sn$_{y}$ which includes a stanene layer sandwiched between Fe$_3$Sn bilayers, shown in Fig. \ref{fig1}h.  Also shown is a corresponding transmission electron microscopy (10$\bar{1}$0)  cross-section of a Fe$_{3}$Sn$_{2}$ single crystal revealing the Fe$_{3}$Sn and stanene layers.  Previous studies have identified Fe$_{3}$Sn$_{2}$ as an unusual magnetic conductor with a high Curie temperature $T_{C} = 670$ K \cite{AllFeSn, Fe3Sn2M}.  While originally attention was focused on the zero field spin structure \cite{Mossbauer}, more recent studies have focused on the system as a potential host for Skyrmion bubbles \cite{fsSkyrmions} and a significant anomalous Hall effect \cite{Fe3Sn2Hall, Fe3Sn2PRB} at finite field.  The latter is particularly interesting in comparison with the related antiferromagnets Mn$_{3}$Sn and Mn$_{3}$Ge recently reported to have a large room temperature anomalous Hall response \cite{Mn3Sn, Felser} and possible Weyl fermion states \cite{Kuroda}.
 
\begin{figure}
\includegraphics[width =  \columnwidth]{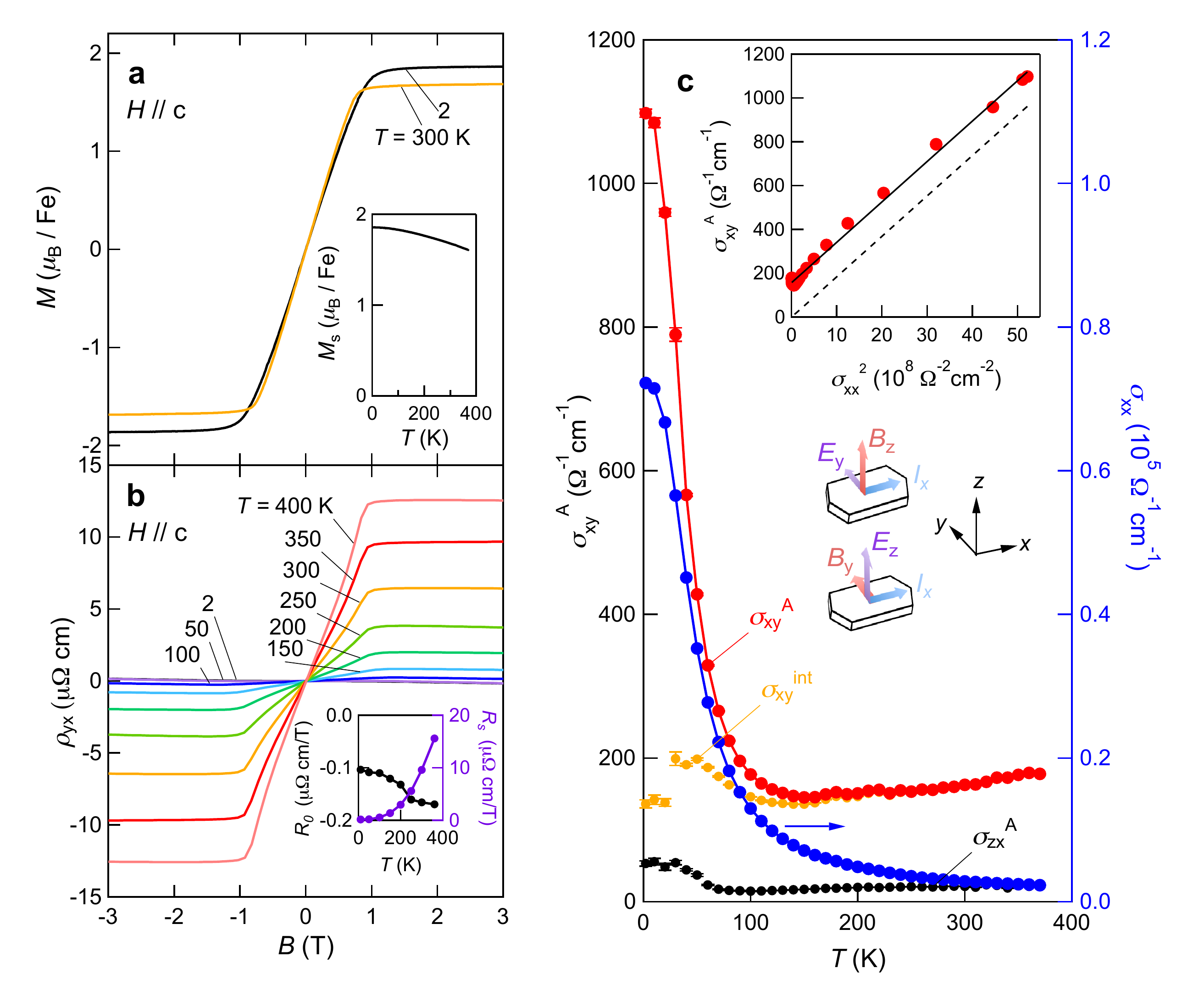}
\caption{\label{fig2}  Ye \emph{et al.}}
\end{figure}

As shown in Fig. \ref{fig2}a, measurements of magnetization $M$ as a function of magnetic induction $B$ along the $c$-axis show that the system is a soft ferromagnet with a mild temperature dependence of saturation field and magnetization $M_{S}$.  The latter reaches approximately 1.9$\mu_{B}$/Fe at low temperature (shown Fig. \ref{fig2}a inset). The crystals show high metallicity with the residual resistivity ratio $\rho(300$ K$)$/$\rho(2$ K$)=25$ (see SM) allowing characterization by electrical transport.  The transverse resistivity in the kagome plane $\rho_{yx}(B)$ (Fig. \ref{fig2}b) strongly reflects $M(H)$, a characteristic of the anomalous Hall effect \cite{Fe3Sn2Hall, Fe3Sn2PRB}.  In ferromagnetic conductors it is conventional to express $\rho_{yx}(B)$ as contributions from the ordinary (Lorentz force) Hall coefficient $R_{0}$ and anomalous Hall coefficient $R_{S}$, \textit{viz}. $\rho_{yx}= R_{0}B +R_{s}M$ \cite{AHEreview}; as shown in the inset of Fig. \ref{fig2}b, $R_{0}$ is mildly $T$ dependent (corresponding to $6\times 10^{21}$ e$^{-}$ cm$^{-3}$ at low $T$) while $R_{S}$ is significantly larger but shows a reduction with lowering $T$.

To further elucidate the role of the bilayer kagome lattice in determining these properties, we examine the associated Hall conductivities in the kagome plane.  The contributions to the total Hall conductivity $\sigma_{xy} = \sigma_{xy}^{N} + \sigma_{xy}^{A}$ ($N(A)$ denotes the normal (anomalous) component) can be separated using the field linearity of $\sigma_{xy}^{N}$ in the low Hall angle limit (see SM).  While $\sigma_{xy}^{A}$ is known to have contributions of both intrinsic (Berry curvature) and extrinsic (scattering) origin \cite{AHEreview}, it has recently been demonstrated that the insensitivity of the latter to thermal excitations allows the parametrization $\sigma_{xy}^A =f(\sigma_{xx,0}) \sigma_{xx}^2 + \sigma_{xy}^{int}$, where $f(\sigma_{xx,0})$ is a function of the residual conductivity $\sigma_{xx,0}$, $\sigma_{xx}$ is the conductivity, and $\sigma_{xy}^{int}$ is the intrinsic anomalous Hall conductivity \cite{Scaling}.  As $\sigma_{xy}^{int}$ does not depend on the scattering rate, in a system with a significant Berry curvature $\sigma_{xy}^{int}$ is then the remnant $\sigma_{xy}^{A}$ observed as $\sigma_{xx}^2 \rightarrow 0$ (shown here inset Fig. \ref{fig2}c).  Fig. \ref{fig2}c shows that at high temperature $\sigma_{xy}^{A}(T)$ remains relatively unchanged from this remnant value until near $T=100$ K where an upturn is observed concomitant with increasing $\sigma_{xx}(T)$.  The latter is indicative of the onset of an extrinsic response expected with the longer relaxation time in this range of $\sigma_{xx}$ \cite{onodaScaling}; the subsequent $\sigma_{xx}^2$ scaling of the additional $\sigma_{xy}^{A}$ (inset of Fig. \ref{fig2}c) is also consistent with an extrinsic origin \cite{Scaling, shitade}.  Thus the scattering rate independent value at high $T$ persists with variation near $10\%$ down to $T=2$ K ($158 \pm 16$ $\Omega^{-1}$cm$^{-1}$) corresponding to approximately 0.27 $e^{2}/h$ per kagome bilayer, which we identify as $\sigma_{xy}^{int}$ (Fig. \ref{fig2}c) akin to that expected from a massive Dirac band \cite{shitade}.

The above observations point to a significant Berry phase contribution to the transport response in Fe$_{3}$Sn$_{2}$ in a geometry that samples the kagome planes (shown as the central inset of Fig. \ref{fig2}c).  We have also measured the Hall response perpendicular to the kagome plane $\sigma_{zx}$ (geometry shown in lower inset of Fig. \ref{fig2}c). As shown in Fig. \ref{fig2}c, the out-of-plane signal is significantly smaller, with the ratio $|\sigma_{zx}^{A} / \sigma_{xy}^A|$ being less than 10$\%$ at the highest $T$, indicating a significant relative enhancement of the Berry curvature in the kagome plane.  We see a consistently enhanced in-plane response across different samples (see SM).   

\begin{figure}
\includegraphics[width = \columnwidth]{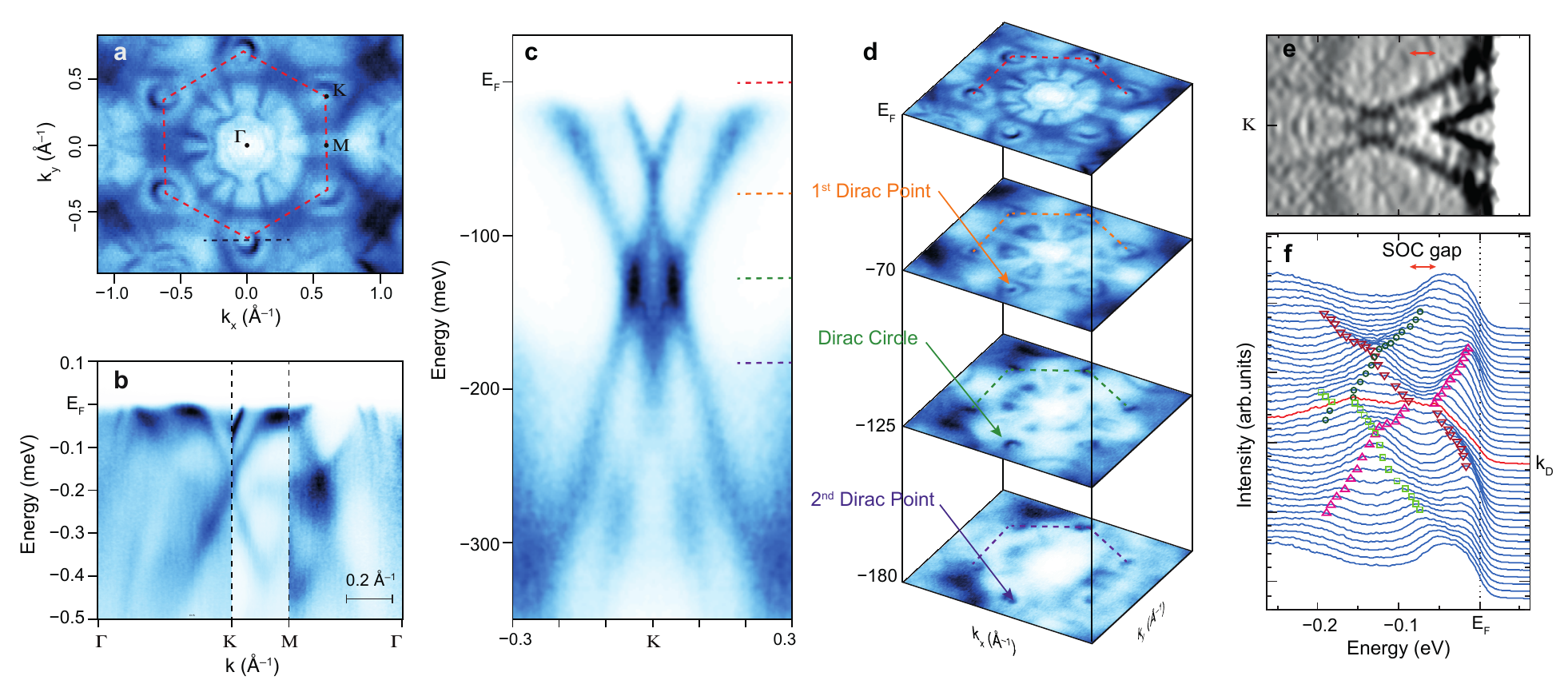}
\caption{\label{fig3}  Ye \emph{et al.}}
\end{figure}

To further examine the origin of this Hall response, we measured the electronic structure of Fe$_{3}$Sn$_{2}$ by angle-resolved photoemission spectroscopy (ARPES). Figs. \ref{fig3}a and \ref{fig3}b show the experimental Fermi surface and energy-momentum dispersion, respectively, of the electronic bands along high symmetry directions parallel to the kagome planes measured at $T = 20$ K (see also Supplementary Fig. 2).  A rich spectrum of electronic excitations with hexagonal symmetry is observed, consistent with the metallicity and crystallographic structure described above.  More notably, linearly dispersing Dirac cones are observed at the Brillouin zone corner points $K$ and $K'$.  As we describe in detail below, this spectrum reminiscent of the electronic structure of graphene \cite{Rotenberg2007} is the long-sought realization of kagome-derived Dirac fermions.  These Dirac-like bands are shown in detail in the high resolution energy-momentum section of the ARPES data across the $K$ point in Fig. \ref{fig3}c (data are taken along the blue dashed line in Fig. \ref{fig3}a, and then symmetrized in momentum about $K$); two Dirac cones separated in energy but both centered at $K$ are resolved.  Hereafter, we focus on these bands and their subsequent role in generating Berry curvature.

The two-fold Dirac cones can be similarly identified in the constant energy contours as shown in Fig. \ref{fig3}d. At the Fermi energy (top layer in Fig. \ref{fig3}d), a pair of Dirac cones forms two electron pockets centered at the $K$ point: a circular inner pocket and a trigonally-warped outer pocket. Moving down from the Fermi energy each pocket shrinks, forming apparent Dirac points at binding energies of 70 meV (second layer from top in Fig. \ref{fig3}d) and 180 meV (bottom layer in Fig. \ref{fig3}d). At the midpoint energy (125 meV), the two Dirac cones cross and, within our experimental resolution, form a ring of Dirac points in the $xy$ momentum plane. The experimental electronic structure of Fe$_{3}$Sn$_{2}$ near the $K$ point is thus characterized by two energy-split ($\Delta E = 110$ meV) interpenetrating Dirac cones.  As we describe below, this splitting is a natural consequence of the bilayer kagome structure similar to the case of multilayer graphene \cite{Rotenberg2013}, whereas the exchange splitting due to magnetic order is expected to be a significantly larger energy scale (in excess of 2 eV \cite{Fe3Sn}).  Photon-energy-dependent ARPES (Supplementary Fig. 3) reveals a negligible variation of the Dirac bands as a function of out-of-plane momentum $k_{z}$, indicating quasi-two-dimensional bands confined to the Fe kagome bilayer. The velocity of the Dirac fermions $v_{D}$ is found to be isotropic in the kagome plane with magnitude $v_{D}=(1.76\pm0.11)\times10^{5} $ m$/$s, comparable to that observed recently in iron pnictide \cite{BaFe2As2} and selenide \cite{FeSe} superconductors, a possible reflection of the correlated character of the Fe-3$d$ states.

Having established the Dirac fermiology of Fe$_{3}$Sn$_{2}$, we focus on the role of spin-orbit coupling and possible mass acquisition of the Dirac bands (see again Fig. \ref{fig1}f). The Dirac band centered at 70 meV can be reliably analyzed, whereas matrix element effects and the proximity of neighboring bands interfere with the intensity distribution of the lower Dirac point.  Inspection of the raw ARPES data shows significantly suppressed spectral intensity at the Dirac point (see supplementary Fig. 3d) which is more clearly visualized in the second derivative of the ARPES map shown in Fig. \ref{fig3}e.  Analysis of the energy distribution curves (EDCs) displayed in Fig. \ref{fig3}f reveals a break between the upper and lower branches of the Dirac cone, signaling the opening of an energy gap $\Delta$. A quantitative analysis performed by fitting the averaged EDCs with Lorentzian peaks returns $\Delta = 30 \pm 5$  meV.  This value is of similar order as that predicted previously for spin-orbit coupled 3$d$ transition metals in a kagome lattice \cite{kagomeCI}.  Compared to previous reports of massive Dirac bands in ARPES, $\Delta$ here is intermediate to the case of smaller gaps in magnetically doped topological insulators \cite{ZX2010, Hasan2012} and larger masses induced in graphene \cite{Hoffman2010, Conrad}. 

\begin{figure}
\includegraphics[width = \columnwidth]{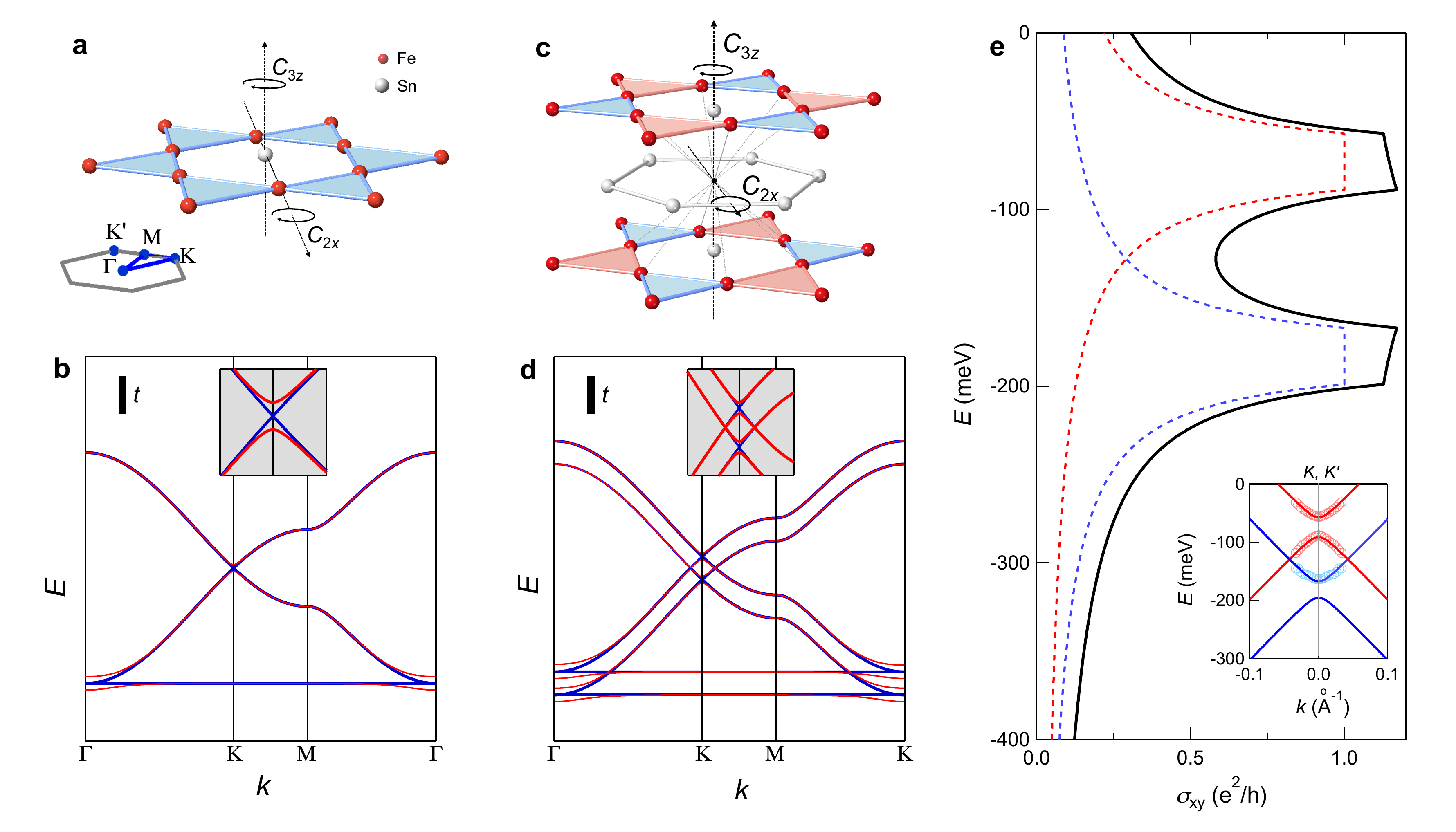}
\caption{\label{fig4} Ye \emph{et al.}}
\end{figure}

The emergence of massive Dirac fermions in Fe$_{3}$Sn$_{2}$ can be understood as a combination of ferromagnetic splitting and spin-orbit coupling in the underlying kagome geometry.  Motivated by the weak $k_{z}$ dispersion observed in ARPES, we consider a stacked system of kagome layers.  Fig \ref{fig4}a shows a perfect Fe$_{3}$Sn kagome layer and the corresponding Brillouin zone. The kagome layer has two-fold and three-fold rotational symmetries ($\mathcal{C}_{2x}$ and $\mathcal{C}_{3z}$, respectively) that leave the $K$ and $K'$ points invariant and thus form point group $\mathcal{D}_{3}$.  In the absence of spin-orbit coupling, the two-fold degenerate crossing (Dirac) point at the Brillouin zone boundary $K$ and $K'$ belong to a two-fold $E$ representation and hence are protected. As shown in Fig. \ref{fig4}b, a Dirac crossing can be observed at $K$ in a tight binding model for nearest neighbor hopping on the kagome sites $\mathcal{H}_{K} = \sum_{<ij>}tc^{\dagger}_{i}c_{j}$, where $<ij>$ indexes nearest-neighbor pairs, $t$ is the hopping integral and $c$ ($c^{\dagger}$) is the fermion annihilation (creation) operator taken to be spin-polarized due to exchange.  The kagome bilayers in Fe$_{3}$Sn$_{2}$ shown in Fig. \ref{fig4}c are tiled by triangles of two different bond lengths, 2.59 \AA{} and 2.75 \AA,  as indicated by the red and blue shading.  However, the combined unit of these kagome layers and the intervening stanene layer preserves the $\left\{\mathcal{C}_{2x}, \mathcal{C}_{3z}\right\}$ symmetry of the perfect kagome lattice and thus the Dirac points are protected by crystal symmetry in the absence of spin-orbit coupling. The additional layer degree of freedom further enriches the electronic structure.  In particular, the ABA layer stacking of the structure in Fig. \ref{fig4}c gives rise to bonding-antibonding splitting \cite{Hiroi}, as seen in a simple tight binding model with this additional hopping (Fig. \ref{fig4}d).

We next introduce Kane-Mele-type spin-orbit coupling $\mathcal{H}_{\text{SOI}} = i\sum_{<ij>}\lambda_{ij}\left[c^{\dagger}_{i, \uparrow}c_{j, \uparrow} - c^{\dagger}_{i, \downarrow}c_{j, \downarrow}\right]$ to the tight binding model $\mathcal{H}_{K}$, where $\lambda_{ij}$ represents the effect of spin-orbit coupling \cite{KaneMele}.  Writing $\lambda_{ij} = \lambda [\vv{E}_{ij} \times \vv{R}_{ij}] \cdot \vv{s}$  ($\lambda$ is the spin-orbit coupling constant, $\vv{E}$ is the electric field on the hopping path, $\vv{R}$ is the hopping vector, and $\vv{s}$ the Pauli matrices of the electron spin) for spin-polarized bands near the $K$ and $K'$ point with finite $s_{z}$ polarization, the $\mathcal{H}_{\text{SOI}}$ effectively reduces to the Haldane term \cite{HaldaneQHE}.  Accordingly, for the single layer case (Fig. \ref{fig4}b) when the Fermi level is positioned in the Dirac gap, the system enters a Chern insulating phase with quantized anomalous Hall conductance \cite{HaldaneQHE, kagomeCI}.  

To connect with the Hall response, we can construct a $k\cdot p$ Hamiltonian near $K$ and $K'$ for the dual Dirac fermions:
\begin{equation}
H_{D}=[\hbar v_F(k_x\sigma_y-k_y\sigma_x)]\otimes I +E_{0}\tau_x+ m\sigma_z,
\label{D}
\end{equation}
where $\vv{\sigma}$ are the Pauli matrices of pseudospin for each Dirac band, $E_{0}$ the energy splitting of the Dirac bands described by the Pauli matrix $\tau_{x}$, and $m = \Delta / 2$ is the Dirac mass.  To obtain the band parameters, we fit the observed dispersion $E(k)$ to the massive Dirac model $E_{\pm}^{i}(k) = \pm \sqrt{(\hbar k v_{D})^{2}+(\Delta / 2)^{2}} + E_{0}^{i}$, where $E_{0}^{i}$ is the energy offset for the $i=1(2)$ upper (lower) Dirac band from $E_{F}$.  As shown in the inset of Fig. \ref{fig4}e, a satisfactory fit is found with $v_{D} = 1.85 \times 10^{5}$ m$/$s, $\Delta$ = 32 meV, and $E_{0}^{1(2)} = -73 (-182)$ meV, comparable to the values extracted above.  Applying these for both upper and lower Dirac bands, we then calculate the contribution of the massive Dirac bands to the Hall response by integrating the Berry curvature over the filled states described by Eq. (\ref{D}) (see SM).  Importantly, the massive Dirac fermions at the $K$ and $K'$ valleys are related by inversion symmetry and therefore contribute similarly to the Berry curvature. As shown in Fig. \ref{fig4}e, the compounded contributions of the Dirac gaps to the Hall conductivity evaluates to $\sigma_{xy}^{calc} = 0.31$ $e^{2}/h$ at $E_{F}$ for a kagome bilayer, comparable to the observed value $\sigma_{xy}^{int} = 0.27$ $e^{2}/h$ per bilayer.  Remarkably, despite the complex fermiology of Fe$_{3}$Sn$_{2}$, the action of the massive Dirac fermions at $K$ and $K'$ largely accounts for the observed Hall response viewed as a simple parallel network of bilayer kagome planes.  This is likely due to the concentration of Berry curvature in the quasi-2D massive Dirac bands which have a small $E_{F}$ comparable to the spin-orbit coupling strength \cite{Nagaosa}.  The robustness of this effect is comparable to that driven by chiral antiferromagnetic order in Mn$_{3}$Sn \cite{Mn3Sn} and Mn$_{3}$Ge \cite{Felser}, but instead of 3D Weyl nodes \cite{Kuroda} is driven by quasi-2D Dirac fermions borne out of a simple ferromagnetic kagome network interleaved with stanene layers.

By combining transport, ARPES, and theoretical analysis, the present study provides a comprehensive proof of principle for the engineering of band structure singularities and topological phenomena in correlated systems. In particular, it demonstrates how the underlying symmetries and lattice geometry can determine the physical properties of a complex electronic system.  We suggest the transport and spectroscopic signatures presented here identify the essential properties of a ferromagnetic kagome metal and that a number of materials in this and structurally related classes are potential intrinsic platforms for correlated topological phases.  Viewed in isolation, the bands near $K$ can be considered to exhibit a two-dimensional ``Chern gap'' - a time-reversal symmetry broken topologically non-trivial phase- intrinsic to a stoichiometric material which has a dominant contribution to the electrical response reaching $T > 300$ K.  This indicates the possibility of engineering such stoichiometric materials down to atomically thin layers to realize the quantum anomalous Hall effect with orders of magnitude larger energy scales than presently possible \cite{QAH} towards the realization of room temperature dissipationless modes.  Extending this approach to stabilize the dispersionless states similarly expected from the kagome network adds an exciting prospect for increasing correlations and enabling the study of magnetically driven fractionalization of states \cite{kagomeFCI, kagomeTSM}.

\section{Methods}

\textbf{Single Crystal Growth} Single crystals of Fe$_{3}$Sn$_{2}$ were grown using an I$_{2}$ catalyzed reaction. Stoichiometric ratio of Fe and Sn powders are sealed in a quartz tube with approximately 1\% I$_2$ by mass and kept in a horizontal 3-zone furnace with a temperature gradient from 750 $^{\circ}$C to 650 $^{\circ}$C for 5 weeks followed by water quenching to stabilize the Fe$_3$Sn$_2$ phase. Hexagonal, plate-like crystals of sub-mm size (see the inset of Fig. S1a) form near the high temperature region as has been previously reported for Fe$_3$Ge \cite{Fe3Ge}.  The hexagonal surfaces are confirmed as (0001) kagome planes by single crystal X-ray diffraction.

\textbf{Magnetization Measurements} Magnetization measurements are performed using a commercial SQUID magnetometer with the field oriented along the $c$-axis and in the $ab$-plane.  Demagnetization corrections were performed for all measurements.

\textbf{Transport Measurements} Four probe transport measurements are performed for longitudinal and Hall resistivity in a commercial cryostat with a superconducting magnet.  High field transport measurements in fields up to 31 T were performed in a He-3 cryostat at Cell-9 of the National High Magnetic Field Laboratory.  For measurements in the kagome plane, the field is applied along the [001] direction with current and voltages in the kagome plane.  For Hall measurements perpendicular to the kagome plane, the magnetic field and current are applied orthogonally in the kagome plane and the out plane voltage is measured.  Electrical signals are detected using standard AC lock-in techniques with a typical current density 10 A/cm$^2$.  To correct for contact misalignment, the measured longitudinal and transverse voltages were field symmetrized and antisymmetrized, respectively.  Demagnetization corrections were performed for all measurements.  

\textbf{Scanning Transmission Electron Microscopy (STEM)}
STEM experiments were conducted at a probe corrected JEOL  ARM STEM  operated at 200 kV acceleration voltage. Fe$_{3}$Sn$_{2}$ samples were prepared by a FEI Helios Focused Ion Beam, operated at 30 kV acceleration voltage for the Gallium beam during lift-out and 2 kV during polishing. Additional polishing was performed at 1 kV and 0.5 kV with a Fischione NanoMill. At both acceleration voltages, samples were polished for 20 min on each side.

\textbf{Angle-resolved photoemission spectroscopy (ARPES)} ARPES experiments were conducted at Beamline 7 (main data) and Beamline 4 (preliminary measurement) of the Advanced Light Source. The two ARPES endstations are equipped with R4000 hemispherical electron analyzers (VG scienta, Sweden). Data in Fig. 3 and Fig. S3 were collected at 20 K with the photon energy of 92 eV, which maximizes the ARPES spectral weight of the Dirac bands. The photon energy dependent measurement is conducted from 45 eV to 120 eV. Energy and momentum resolutions were better than 15 meV and 0.01 \AA$^{-1}$, respectively. Fe$_{3}$Sn$_{2}$ samples were cleaved in the ultrahigh vacuum chamber with the base pressure better than 4$\times$10$^{-11}$ Torr.  All the data are collected within 8 hours after cleaving to minimize the effect of sample degradation. Six different samples from various growth batches were analyzed to confirm consistency of results.

\newpage
\textbf{Acknowledgments}
We are grateful to X.-G. Wen and E. Tang for fruitful discussions. This research is funded in part by the Gordon and Betty Moore Foundation EPiQS Initiative, Grant GBMF3848 to JGC and NSF grant DMR-1554891.  L.Y., J.L., and F.v.C. acknowledge support by the STC Center for Integrated Quantum Materials, NSF Grant No. DMR-1231319.  L.Y. acknowledges support by the Tsinghua Education Foundation.  M.K. acknowledges a Samsung Scholarship from the Samsung Foundation of Culture.  This research used resources of the Advanced Light Source, which is a DOE Office of Science User Facility under contract no. DE-AC02-05CH11231.  A portion of this work was performed at the National High Magnetic Field Laboratory, which is supported by National Science Foundation Cooperative Agreement No. DMR-1157490, the State of Florida, and the US Department of Energy.

\hfill

\newpage

\textbf{Figure 1. The kagome structure and Fe$_{3}$Sn$_{2}$}  \textbf{a} Structure of kagome lattice and \textbf{b} associated Dirac point in the nearest neighbor tight-binding model. \textbf{c} Ferromagnetic kagome lattice with \textbf{d} associated spin-polarized Dirac band. \textbf{e} Spin-orbit coupled ferromagnetic kagome lattice with Berry phase $\phi$ accrual via hopping and \textbf{f} corresponding gapped Dirac spectrum.  \textbf{g}  The Fe$_{3}$Sn kagome plane in Fe$_{3}$Sn$_{2}$ with kagome network shown in red. \textbf{h} Transmission electron microscopy cross-section of Fe$_{3}$Sn$_{2}$ and corresponding Fe$_{3}$Sn and stanene layers viewed from the [$10\bar{1}0$] direction. 
  \\

\textbf{Figure 2. Anomalous Hall response of Fe$_{3}$Sn$_{2}$} \textbf{a} Magnetization of Fe$_{3}$Sn$_{2}$ along the $c$-axis as a function of magnetic induction $M(B)$ at room temperature ($T=300$ K) and low temperature ($T=2$ K). The inset shows the saturation magnetization (measured at 2 T)  as a function of $T$. \textbf{b} Hall resistivity $\rho_{yx}$ as a function of $B$. The inset shows the ordinary and anomalous Hall coefficients $R_0$ and $R_{S}$, respectively, as a function of $T$. \textbf{c} Anomalous Hall conductivities $\sigma_{xy(zx)}^{A}$ in the $xy$ ($zx$)-plane along the longitudinal conductivity $\sigma_{xx}$ and estimated intrinsic Hall conductivity $\sigma_{xy}^{int}$.  The measurement configurations for both  $\sigma_{xy}$ and $\sigma_{zx}$ are shown in the lower inset. The upper inset shows $\sigma_{xy}^A$ plotted versus $\sigma_{xx}^2$; the solid and dashed lines are the scaling curves (see text). 
 \\

\textbf {Figure 3. Massive Dirac fermion at the zone corner of Fe$_{3}$Sn$_{2}$} \textbf{a} Experimental Fermi surface of Fe$_{3}$Sn$_{2}$. The hexagonal Brillouin zone and high symmetry points are marked with red dashed line and black dots. \textbf{b} Experimental band dispersion of Fe$_{3}$Sn$_{2}$ along the high symmetry directions. \textbf{c} High resolution ARPES data taken along the blue dashed line in \textbf{a}, and subsequently symmetrized with respect to $K$. A pair of Dirac cones separated in energy axis is clearly visible. \textbf{d} Constant energy maps at the binding energy of 0, 70, 125, and 180 meV. Two electron pockets, the first Dirac point, the Dirac circle, and the second Dirac point can be clearly detected from respective maps. \textbf{e},\textbf{f} The second derivative plot \textbf{e} and the stack of EDCs \textbf{f} across the Dirac points. Both panels share the momentum range and direction with \textbf{c}. The red double-headed arrow marks the discontinuity between the upper and lower branches of the Dirac cone. All data are taken with 92 eV photons.\\

\textbf{Figure 4. Tight Binding and Hall Conductivity of a Kagome Bilayer}   \textbf{a} Two-fold ($\mathcal{C}_{2x}$) and three-fold ($\mathcal{C}_{3z}$) rotation axis symmetry operations of a single Fe$_{3}$Sn kagome layer. \textbf{b} Tight binding band model of single layer kagome with (red) and without (blue) spin orbit coupling.  The inset shows a magnified of the (avoided) crossing near $K$.  \textbf{c} $\mathcal{C}_{2x}$ and $\mathcal{C}_{3z}$ symmetries of the breathing kagome and stanene layers. \textbf{d} Tight binding of double layer kagome with in-plane (interplane) hopping $t$ ($0.3t$) and with (red) and without (blue) spin-orbit coupling.  The upper inset shows a magnified view of the double Dirac structure near $K$. The spin orbit coupling strength $\lambda=0.05$ for both \textbf{b} and \textbf{e}. \textbf{e} Anomalous Hall conductivity as a function of Fermi energy from the $k \cdot p$ model for the ferromagnetic kagome bilayer (see text).  The red (blue) curve represents the contribution from the upper (lower) Dirac band.  The inset shows the fit of a massive Dirac dispersion to the ARPES results near $K$ to determine the parameters for the $k \cdot p$ theory.\\

\clearpage

\clearpage

\clearpage

\clearpage

\begin{thebibliography}{99}

\bibitem{CrystalStructure} O'Keeffe, M. \& Hyde, B. G. \emph{Crystal structures, I. Patterns and symmetry}, (Mineralogical Society of America, 1996).
\bibitem{KagomeTheory} Sachdev, S. Kagome- and triangular-lattice Heisenberg antiferromagnets: Ordering from quantum fluctuations and quantum-disordered ground states with unconfined bosonic spinons. \emph{Phys. Rev. B} \textbf{45}, 12377 (1992).
\bibitem{jarosite} Inami, T., Nishiyama, M., Maegawa, S. \& Oka, Y. Magnetic structure of the kagome lattice antiferromagnet potassium jarosite KFe$_3$(OH)$_6$(SO$_4$)$_2$. \emph{Phys. Rev. B} \textbf{61}, 12181-12186 (2000).
\bibitem{volborthite} Hiroi, Z., Hanawa, M., Kobayashi, N., Nohara, M., Takagi, H., Kato, Y. \& Takigawa, M. Spin-1/2 kagom\'{e}-Like lattice in Volborthite Cu$_3$V$_2$O$_7$(OH)$_2$·2H$_2$O. \emph{J. Phys. Soc. Jpn.} \textbf{70}, 3377-3384 (2001).
\bibitem{CuBDC} Chisnell, R., Helton, J. S., Freedman, D. E., Singh, D. K., Bewley, R. I., Nocera, D. G., \& Lee, Y. S. Topological magnon bands in a kagome lattice ferromagnet. \emph{Phys. Rev. Lett.} \textbf{115}, 147201 (2015).
\bibitem{herber}  Han, T. -H., \emph{et al.}, Fractionalized excitations in the spin-liquid state of a kagome-lattice antiferromagnet. \emph{Nature} \textbf{492}, 406-410 (2012).
\bibitem{QSL-RMP} Zhou, Y., Kanoda, K., \& Ng, T. -K.  Quantum spin liquid states. \emph{Rev. Mod. Phys.} \textbf{89}, 025003 (2017).
\bibitem{Mazin} Mazin,I. I.\emph{et al.} Theoretical prediction of a strongly correlated Dirac metal. \emph{Nat. Comm.} \textbf{5}, 4261 (2014).
\bibitem{kagomeTI}  Guo, H.-M. \& Franz, M. Topological insulator on the kagome lattice. \emph{Phys. Rev. B} \textbf{80}, 113102 (2009)
\bibitem{kagomeCI} Xu, G., Lian, B., \& Zhang, S.-C. Intrinsic quantum anomalous Hall effect in the kagome lattice Cs$_2$LiMn$_3$F$_{12}$. \emph{Phys. Rev. Lett.} \textbf{115}, 186802 (2015).
\bibitem{Mn3Sn} Nakatsuji, S., Kiyohara, N. \& Higo, T. Large anomalous Hall effect in a non-collinear antiferromagnet at room temperature. \emph{Nature} \textbf{527}, 212-215 (2015).
\bibitem{Felser} Nayak, A. K. \emph{et al}. Large anomalous Hall effect driven by a nonvanishing Berry curvature in the noncolinear antiferromagnet Mn$_{3}$Ge.  \emph{Science Advances} \textbf{2}, 4 (2016).
\bibitem{Kuroda} Kuroda, K. \emph{et al}.  Evidence for magnetic Weyl fermions in a correlated metal.  \emph{Nat. Mater.}  doi:10.1038/nmat4987 (2017).
\bibitem{kagomeFCI} Tang, E. \& Wen, X. -G. High-Temperature Fractional Quantum Hall States. \emph{Phys. Rev. Lett.} \textbf{106}, 236802 (2011).
\bibitem{kagomeTSM} Bergholtz,E. J.,  Liu, Z., Trescher, M. Moessner, R. \& Udagawa, M. Topology and interactions in a frustrated slab: tuning from Weyl semimetals to $\mathcal{C}>1$ fractional Chern insulators. \emph{Phys. Rev. Lett.} \textbf{114}, 016806 (2015).
\bibitem{flatbandF} Tasaki, H. From Nagaoka's ferromagnetism to flat-band ferromagnetism and beyond: An introduction to ferromagnetism in the Hubbard model. \emph{Prog. Theo. Phys.} \textbf{99}, 489-548 (1998).
\bibitem{Superconductivity} Bauer, B.\emph{et al.}, Chiral spin liquid and emergent anyons in a kagome lattice Mott insulator. \emph{Nat. Comm.} \textbf{5}, 5137 (2014).
\bibitem{grapheneband}  Wallace, P. R. The band theory of graphite. \emph{Phys. Rev.} \textbf{71}, 622-634 (1947).
\bibitem{LiangCI} Zhu, W., Gong, S.-S., Zeng, T.-S., Fu, L., \& Sheng, D. S.  Interaction-driven spontaneous quantum Hall effect on a kagome lattice.  \emph{Phys. Rev. Lett.} \textbf{117}, 096402 (2016).
\bibitem{AHEHaldane} Haldane, F. D. M. Berry curvature on the Fermi surface: anomalous Hall effect as a topological Fermi-liquid property. \emph{Phys. Rev. Lett.} \textbf{93}, 206602 (2004).
\bibitem{magnonHall} Onose, Y.\emph{et al}, Observation of the magnon Hall effect. \emph{Science}, \textbf{329}, 297-299 (2010).
\bibitem{AllFeSn} Giefers, H. \& Nicol, M. High pressure X-ray diffraction study of all Fe–Sn intermetallic compounds and one Fe–Sn solid solution. \emph{J. Alloy Comp.} \textbf{422}, 132-144 (2006).
\bibitem{Fe3Sn2M} Fenner, L. A., Dee, A. A. \& Wills, A. S., Non-collinearity and spin frustration in the itinerant kagome ferromagnet Fe$_3$Sn$_2$. \emph{J. Phys.: Condens. Matter} \textbf{21}, 452202 (2009)
\bibitem{Mossbauer} Ca$\ddot{\text{e}}$rt,  G. L., Malaman B., \& Roques, B., M$\ddot{\text{o}}$ssbauer effect study of Fe$_3$Sn$_2$. \emph{J. Phys. F: Metal Phys.} \textbf{8}, 323 (1978).
\bibitem{fsSkyrmions} Hou, Z. \textit{et al}., Observation of various and spontaneous magnetic Skyrmionic bubbles at room temperature in a frustrated kagome magnet with uniaxial magnetic anisotropy. \emph{Adv. Mater.}  \textbf{2017}, 1701144 (2017). 
\bibitem{Fe3Sn2Hall} Kida, T.\emph{et al.}, The giant anomalous Hall effect in the ferromagnet Fe$_3$Sn$_2$ - a frustrated kagome metal \emph{J. Phys.: Condens. Matter} \textbf{23}, 112205 (2011).
\bibitem{Fe3Sn2PRB} Wang, Q., Sun, S., Zhang, X., Pang, F., \& Lei, H. Anomalous Hall effect in a ferromagnetic Fe$_3$Sn$_2$ single crystal with a geometrically frustrated Fe bilayer kagome lattice. \emph{Phys. Rev. B} \textbf{94}, 075135 (2016)
\bibitem{AHEreview} Nagaosa, N., Sinova, J., Onoda, S., MacDonald, A. H. \& Ong, N. P. Anomalous Hall effect. \emph{Rev. Mod. Phys.} \textbf{82}, 1539-1592 (2010). 
\bibitem{Scaling} Tian, Y., Ye, L., \& Jin, X. Proper scaling of the anomalous Hall effect. \emph{Phys. Rev. Lett.} \textbf{103}, 087206 (2009).
\bibitem{onodaScaling} Onoda, S., Sugimoto, N., \& Nagaosa, N.  Quantum transport theory of anomalous electric, thermoelectric, and thermal Hall effects in ferromagnets.  \emph{Phys. Rev. B} \textbf{77}, 165103 (2008).
\bibitem{shitade} Shitade, A \& Nagaosa, N.  Anomalous Hall effect in ferromagnetic metals: role of phonons at finite temperature.  \emph{J. Phys. Soc. Jpn.} \textbf{81}, 083704 (2012).
\bibitem{Rotenberg2007} Bostwick, A., Ohta, T., Seyller, T., Horn, K., \& Rotenberg, E. Quasiparticle dynamics in graphene. \emph{Nat. Phys.} \textbf{3}, 36-40 (2007).
\bibitem{Rotenberg2013} Kim, K. S., Walter, A. L.,	Moreschini, L.,	Seyller, T., Horn, K., Rotenberg, E., \& Bostwick, A. Coexisting massive and massless Dirac fermions in symmetry-broken bilayer graphene. \emph{Nat. Mater.} \textbf{12} 887-892 (2013).
\bibitem{Fe3Sn} Sales, B. C., Saparov, B., McGuire, M. A., Singh, D. J., \& Parker, D. S.  Ferromagnetism of Fe$_{3}$Sn and alloys.  \emph{Scientific Reports} \textbf{4}, 7024 (2014).
\bibitem{BaFe2As2} Richard., P. \emph{et al.}  Observation of Dirac cone electronic dispersion in BaFe$_{2}$As$_{2}$.  \emph{Phys. Rev. Lett.} \textbf{104}, 137001 (2010).
\bibitem{FeSe} Tan, S. Y. \emph{et al.}  Observation of Dirac cone band dispersions in FeSe thin films by photoemission spectroscopy.  \emph{Phys. Rev. B} \textbf{93}, 104513 (2016).
\bibitem{ZX2010} Chen, Y. L. \emph{et al.} Massive Dirac fermion on the surface of a magnetically doped topological insulator. \emph{Science} \textbf{329} 659-662 (2010).
\bibitem{Hasan2012} Xu, S. Y. \emph{et al.} Hedgehog spin texture and Berry’s phase tuning in a magnetic topological insulator. \emph{Nat. Phys.} \textbf{8} 616-622 (2012).
\bibitem{Hoffman2010} Balog, R. \emph{et al.} Bandgap opening in graphene induced by patterned hydrogen adsorption. \emph{Nat. Mater.} \textbf{9}, 315-319 (2010).
\bibitem{Conrad} Nevius, M. S., Conrad, M., Wang, F., Celis, A., Nair, M. N., Taleb-Ibrahimi, A., Tejeda,  A., \& Conrad, E. H.  Semiconducting graphene from highly ordered substrate interactions.  \emph{Phys. Rev. Lett.} \textbf{115}, 136802 (2015).
\bibitem{Hiroi} Ishii, Y., Harima, H., Okamoto, Y., Yamaura, J., \& Hiroi, Z.  YCr$_{6}$Ge$_{6}$ as a candidate compound for a kagome metal.  \emph{J. Phys. Soc. Jpn.}
\textbf{82}, 023705 (2013).
\bibitem{KaneMele} Kane, C. L. \& Mele, E. J.  Quantum Spin Hall Effect in Graphene.  \emph{Phys. Rev. Lett.} \textbf{95}, 226801 (2005).
\bibitem{HaldaneQHE} Haldane, F. D. M.  Model for a Quantum Hall Effect without Landau Levels: Condensed-Matter Realization of the ``Parity Anomaly.''  \emph{Phys. Rev. Lett.} \textbf{61}, 2015-2018 (1998).
\bibitem{Nagaosa} Fang, Z., Nagaosa, N., Takahashi, K. S., Asamitsu, A., Mathieu, R., Ogasawara, T., Yamada, H., Kawasaki, M., Tokura, Y., \& Terakura, K.  The Anomalous Hall effect and magnetic monopoles in momentum space.  \emph{Science} \textbf{302}, 92-95 (2003).
\bibitem{QAH} Chang, C.-Z.  \emph{et al.}  Experimental observation of the quantum anomalous Hall effect in a magnetic topological insulator.  \emph{Science} \textbf{340}, 167-170 (2013).
\bibitem{Fe3Ge} Drijver, J. W., Sinnema, S. G., \& van der Woude, F., Magnetic Properties of hexagonal and cubic Fe$_3$Ge. \emph{J. Phys. F: Metal Phys.} \textbf{6}, 2165-2177 (1976).
\end{thebibliography}
\end{document}